\documentclass[]{mn2e}
\usepackage{epsfig}
\def\lsim{\vcenter{\hbox{$<$}\offinterlineskip\hbox{$\sim$}}}

\title[LI-LMC 1813 in the LMC cluster KMHK 1603]{The superwind mass-loss rate
of the metal-poor carbon star LI-LMC 1813 in the LMC cluster KMHK
1603\thanks{Based on observations collected at the European Southern
Observatory, Chile (ESO N$^\circ$56.E-0681, 66.D-0318 \& 68.D-0660)}}
\author[Jacco Th. van Loon et al.]
{Jacco Th. van Loon$^1$,
 Jonathan R. Marshall$^1$,
 \newauthor
 Mikako Matsuura$^2$
 and Albert A. Zijlstra$^2$\\
$^1$Astrophysics Group, School of Chemistry \& Physics, Keele University,
    Staffordshire ST5 5BG, United Kingdom\\
$^2$UMIST, Department of Physics, P.O. Box 88, Manchester M60 1QD, United
    Kingdom}
\date{Accepted ????.
      Received ????;
      in original form ????}
\pagerange{\pageref{firstpage}--\pageref{lastpage}}
\pubyear{????}
\begin{document}
\maketitle
\label{firstpage}
\begin{abstract}
LI-LMC 1813 is a dust-enshrouded Asymptotic Giant Branch (AGB) star, located
in the small open cluster KMHK 1603 near the rim of the Large Magellanic Cloud
(LMC). Optical and infrared photometry between 0.5 and 60 $\mu$m is obtained
to constrain the spectral energy distribution of LI-LMC 1813. Near-infrared
spectra unambiguously show it to be a carbon star. Modelling with the
radiation transfer code Dusty yields accurate values for the bolometric
luminosity, $L=1.5\times10^4$ L$_\odot$, and mass-loss rate,
$\dot{M}=3.7(\pm1.2)\times10^{-5}$ M$_\odot$ yr$^{-1}$. On the basis of
colour-magnitude diagrams, the age of the cluster KMHK 1603 is estimated to be
$t=0.9$--1.0 Gyr, which implies a Zero-Age Main Sequence mass for LI-LMC 1813
of $M_{\rm ZAMS}=2.2\pm0.1$ M$_\odot$. This makes LI-LMC 1813 arguably the
object with the most accurately and reliably determined (circum)stellar
parameters amongst all carbon stars in the superwind phase.
\end{abstract}


\section{Introduction}

In the final stages of their evolution, stars of main-sequence masses $M_{\rm
ZAMS}\sim$\,0.8--8 M$_\odot$ ascend the Asymptotic Giant Branch (AGB). The
end-product of their evolution is a White Dwarf, with a typical mass of
$M_{\rm WD}\sim$\,0.5 M$_\odot$. It is thus clear that near the tip of the
AGB, stars must be losing most of their mass. The rate at which this occurs is
estimated to be $\dot{M}\sim10^{-6}$ to $10^{-4}$ M$_\odot$ yr$^{-1}$, much
faster than the nuclear burning mass consumption rate which normally
determines the rate of stellar evolution (van Loon et al.\ 1999b). Hence the
mass loss is important for the subsequent evolution of the star, as well as
for the chemical enrichment of the inter-stellar medium. Yet our knowledge and
understanding of the mass-loss mechanism and the composition of the mass lost
is incomplete.

Amongst the key factors in supporting a dense outflow from an AGB star are the
low effective temperature, which allows for the formation of circum-stellar
dust, and the high luminosity, which, through radiation pressure on the dust
grains and collissional coupling between the dust and the gas, drives a
stellar wind. Because the dust only forms at several stellar radii from the
star, there has to be an additional mechanism to bring the gas into the dust
formation region --- possibly provided by strong radial pulsations (e.g.\
Bowen \& Willson 1991).

Dynamical models for the circumstellar envelopes of AGB stars suggest that the
mass-loss rate depends strongly on the stellar effective temperature, about
$\dot{M}{\propto}T_{\rm eff}^{-8}$ (Arndt, Fleischer \& Sedlmayr 1997).
Although this seems to be confirmed for the onset of the dusty mass-loss epoch
(Alard et al.\ 2001), it is not clear whether such strong temperature
dependence is also valid during the so-called ``superwind'' phase where the
AGB star is obscured by its optically thick circum-stellar envelope (van Loon
et al.\ 1999b).

A dramatic change in the chemical composition of the mass lost occurs when an
AGB star becomes a carbon star (Iben \& Renzini 1983): during the temporal
ignition of a helium-burning shell, situated in between the degenerate
carbon/oxygen core and the hydrogen-burning shell, the convective mantle may
reach the nuclear burning layer and transport nuclear burning products to the
stellar surface ($3^{\rm rd}$ dredge-up). These thermal pulses (TP) occur on
timescales of $t_{\rm TP}\sim10^4$--$10^5$ yr, and last only briefly. But over
the course of TP-AGB evolution, the surface abundance of carbon may exceed
that of oxygen, forming a carbon star. Due to the strong binding of the CO
molecule, only the remaining oxygen or carbon is used to synthesize other
molecules that nucleate to form dust grains in the circum-stellar envelope.
Hence the circum-stellar chemistry of carbon stars is fundamentally different
from that of oxygen-rich AGB stars.

There is a threshold mass above which $3^{\rm rd}$ dredge-up is strong enough
to make a carbon star, roughly at $M_{\rm ZAMS}>1.2$ M$_\odot$ and mildly
dependent on metallicity (Marigo, Girardi \& Bressan 1999). However, in
massive AGB stars, about $M_{\rm ZAMS}>4$ M$_\odot$, nuclear burning occurs
within the bottom layers of the convective mantle, converting the carbon into
nitrogen: Hot Bottom Burning (HBB). At the very end of AGB evolution, the
diminished mantle mass as a result of prolonged mass loss causes HBB to cease,
and a final thermal pulse may turn the star into a carbon star after all
(Frost et al.\ 1998). These competing processes are likely to depend on the
initial metallicity of the star.

Optical studies of the AGB population in the Large Magellanic Cloud (LMC)
provide support for the $3^{\rm rd}$ dredge-up and HBB to govern the formation
of carbon stars (e.g.\ Groenewegen \& de Jong 1993; Marigo et al.\ 1999).
Recent infrared (IR) studies could assess the effects of mass loss and
possible observational bias in optical studies. These studies confirm the
delayed formation of massive carbon stars after cessation of HBB, but also
suggest that amongst AGB stars in the superwind phase oxygen-rich stars exist
with masses below the threshold for HBB (Zijlstra et al.\ 1996; van Loon et
al.\ 1997, 1998, 1999b; van Loon, Zijlstra \& Groenewegen 1999a).

Much of the uncertainty with respect to the progenitor mass and initial
metallicity of an individual AGB star can be alleviated if the star is a
member of a co-eval stellar cluster. Optical studies provide a great number
of such examples in LMC and SMC clusters, confirming the mass-range for
optically bright carbon stars between $M_{\rm ZAMS, min}\sim$\,1--1.5 and
$M_{\rm ZAMS, max}\sim$\,3--5 M$_\odot$ (Frogel, Mould \& Blanco 1990; Marigo,
Girardi \& Chiosi 1996; van Loon 2002).

Van Loon et al.\ (1998) noted that the oxygen-rich dust-enshrouded OH maser
star IRAS 05298$-$6957 is located in the LMC cluster HS 327, and their
discovery of an optically bright carbon star in the same cluster directly
provided a value for the threshold mass above which HBB occurs: $M_{\rm
ZAMS}=4$ M$_\odot$ (van Loon et al.\ 2001).

Tanab\'{e} et al.\ (1997) found one dust-enshrouded carbon star in each of the
LMC clusters NGC 1783 and NGC 1978, and another one in the SMC cluster NGC
419. They all have moderate luminosities, $5000<L<8000$ L$_\odot$, and
progenitor masses, $M_{\rm ZAMS}\sim$\,1.5--1.6 M$_\odot$. Their mass-loss
rates are $\dot{M}<10^{-5}$ M$_\odot$ yr$^{-1}$, but the relatively long
pulsation periods of $P\sim$\,500 d indicate that they have already shed a
significant fraction of their mass (Nishida et al.\ 2000).

The IRAS point source LI-LMC 1813 (IRAS 06025$-$6712) was identified as a
candidate dust-enshrouded AGB star in the LMC by Loup et al.\ (1997). Its
suspected nature was confirmed after the identification of the extremely red
counterpart (van Loon et al.\ 1997), and it was later found to be coincident
with an LMC cluster (van Loon 1999), KMHK 1603 (Kontizas et al.\ 1990).
Kontizas et al.\ estimate that the number density of LMC star clusters in the
direction of LI-LMC 1813 is $\sim$20 per square degree, with each cluster
covering, on average, less than 1 square arcminute. Hence, in the direction of
LI-LMC 1813, less than 0.6\% of sky is covered by an LMC cluster. There are of
order $10^2$ IRAS-selected AGB stars in the LMC (Loup et al.\ 1997; van Loon
et al.\ 1997), most of which are concentrated towards the LMC bar. One does
not therefore expect any chance coincidence of an IRAS-selected AGB star
within the $<0.6$\% of sky covered by an LMC cluster in this part of the LMC,
rendering LI-LMC 1813 likely to be a physical member of the LMC cluster KMHK
1603.

We present here the results of extensive optical and IR imaging and
spectroscopic follow-up observations of LI-LMC 1813 and the cluster KMHK 1603.
We identify the dust-enshrouded AGB star as a carbon star, and derive accurate
estimates for its progenitor mass, $M_{\rm ZAMS}=2.2$ M$_\odot$, and mass-loss
rate, $\dot{M}=3.7\times10^{-5}$ M$_\odot$ yr$^{-1}$, which we confront with
theoretical models. Remarkably, the mass-loss rate is very high for this
metal-poor ($\sim\frac{1}{5}$ solar) object.

\section{The brightness of LI-LMC 1813}

We have collected optical (Gunn g, r \& i), near-IR (J, K \& L$^\prime$) and
mid-IR (N) imaging data, which we combine with archival IR data (2MASS, Denis,
MSX and IRAS) to describe the spectral energy distribution (SED) of LI-LMC
1813. These data are described in full in Appendix A1.

\subsection{Images}

%
%
\begin{figure}
\psfig{figure=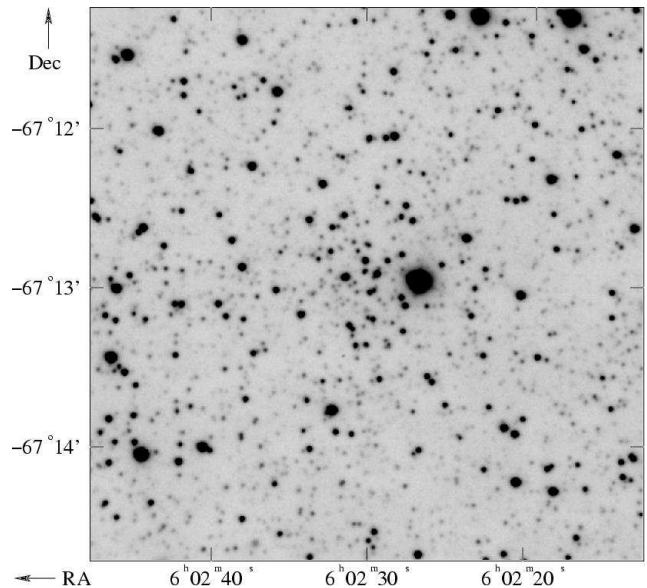,width=84mm}
\caption[]{Image of KMHK 1603 and surrounding field, taken with the 0.9m Dutch
telescope through the i-band filter. The brightest star near the centre of the
field is KMHK 1603-1, whilst LI-LMC 1813 is invisible at this wavelength.}
\end{figure}

%
%
\begin{figure}
\psfig{figure=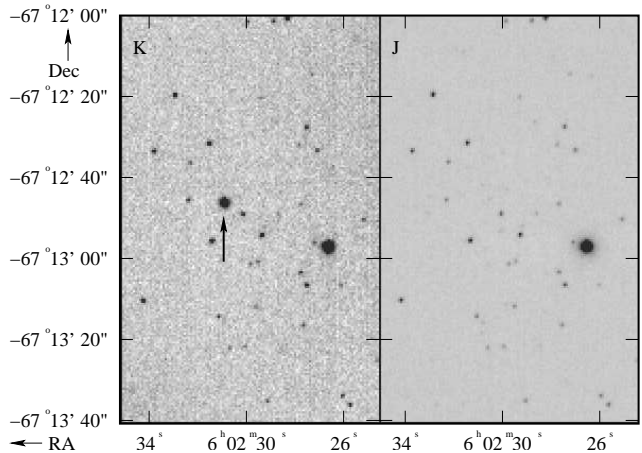,width=84mm}
\caption[]{Near-IR image of KMHK 1603, taken with IRAC-2. The brightest star
in the field, near the edge, is KMHK 1603-1. The very red IR object LI-LMC
1813 is almost equally bright in the K-band (left; arrow) but invisible in the
J-band (right).}
\end{figure}


The moderately crowded optical images (Fig.\ 1) feature the sparse cluster
KMHK 1603, dominated by the bright star KMHK 1603-1. At these wavelengths,
LI-LMC 1813 is invisible. Undetected in the J-band, the brightness of LI-LMC
1813 in the K-band rivals that of KMHK 1603-1 (Fig.\ 2). These are the only
two stars visible at $\lambda\sim$\,3 $\mu$m, with LI-LMC 1813 by then
outshining KMHK 1603-1. At even longer wavelengths, only LI-LMC 1813 is
detected.

\subsection{Photometry}

%
%
\begin{table*}
\caption[]{Photometry of LI-LMC 1813 and the optically bright star KMHK
1603-1: magnitudes for the optical and near-IR, and flux densities (in Jy) for
the mid-IR ($\lambda>6$ $\mu$m). For Denis, MSX and IRAS no exact epoch is
available.}
\begin{tabular}{lllrclcl}
\hline\hline
Band                            &
$\lambda$                       &
Facility                        &
Epoch\hspace{8mm}               &
\multicolumn{2}{c}{LI-LMC 1813} &
\multicolumn{2}{c}{KMHK 1603-1} \\
                      &
($\mu$m)              &
                      &
                      &
Photometry            &
1-$\sigma$            &
Photometry            &
1-$\sigma$            \\
\hline
$g$                   &
0.515                 &
ESO/Dutch 0.9m        &
27 Dec 1996           &
\llap{$>$2}3.0        &
                      &
\llap{1}3.0\rlap{5}   &
                      \\
$r$                   &
0.670                 &
ESO/Dutch 0.9m        &
27 Dec 1996           &
\llap{$>$2}3.0        &
                      &
\llap{1}3.0\rlap{6}   &
                      \\
$i$                   &
0.797                 &
ESO/Dutch 0.9m        &
27 Dec 1996           &
\llap{$>$2}3.0        &
                      &
\llap{1}2.6\rlap{9}   &
                      \\
                      &
                      &
Denis                 &
1995 --- 1999         &
                      &
                      &
\llap{1}2.5\rlap{0}   &
                      \\
$J$                   &
1.25                  &
ESO/IRAC-2            &
2 Jan 1996            &
\llap{$>$2}0.0        &
                      &
\llap{1}2.0\rlap{7}   &
0.04                  \\
                      &
                      &
Denis                 &
1995 --- 1999         &
                      &
                      &
\llap{1}2.1\rlap{6}   &
                      \\
                      &
                      &
2MASS                 &
26 Oct 1998           &
\llap{$>$1}7.5        &
                      &
\llap{1}2.1\rlap{5}   &
0.03                  \\
$H$                   &
1.65                  &
2MASS                 &
26 Oct 1998           &
\llap{1}5.5\rlap{9}   &
0.15                  &
\llap{1}1.8\rlap{7}   &
0.04                  \\
$K$                   &
2.20                  &
ESO/IRAC-2            &
2 Jan 1996            &
\llap{1}3.2\rlap{6}   &
0.02                  &
\llap{1}1.8\rlap{0}   &
0.02                  \\
                      &
                      &
2MASS                 &
26 Oct 1998           &
\llap{1}2.8\rlap{7}   &
0.04                  &
\llap{1}1.8\rlap{0}   &
0.03                  \\
                      &
                      &
CTIO/OSIRIS           &
6 Jul 2002            &
\llap{1}2.3\rlap{5}   &
0.10                  &
                      &
                      \\
$L^\prime$            &
3.78                  &
ESO/ISAAC             &
19 Oct 2000           &
7.7\rlap{4}           &
0.02                  &
\llap{1}1.7\rlap{5}   &
0.05                  \\
                      &
                      &
ESO/ISAAC             &
14 Dec 2001           &
9.2\rlap{7}           &
0.25                  &
                      &
                      \\
\hline
$A$                   &
8.28                  &
MSX                   &
May 1996 --- Feb 1997 &
0.2\rlap{04}          &
0.013                 &
                      &
                      \\
$N2$                  &
\llap{1}0.6           &
ESO/TIMMI-2           &
19 01 2001            &
0.6\rlap{0}           &
0.05                  &
                      &
                      \\
$F_{12}$              &
\llap{1}2             &
IRAS                  &
1983                  &
0.4\rlap{0}           &
0.03                  &
                      &
                      \\
$F_{25}$              &
\llap{2}5             &
IRAS                  &
1983                  &
0.2\rlap{5}           &
0.02                  &
                      &
                      \\
$F_{60}$              &
\llap{6}0             &
IRAS                  &
1983                  &
0.1\rlap{0}           &
0.04                  &
                      &
                      \\
\hline
\end{tabular}
\end{table*}

The complete set of optical and infrared photometry in the range 0.5--60
$\mu$m for LI-LMC 1813 is listed in Table 1. Also included in the table is
photometry for KMHK 1603-1, whose IRAC-2, 2MASS and Denis magnitudes, which
have been transformed onto the SAAO system (Carter 1990; Carpenter 2001), are
in excellent agreement with each other.

\subsection{Variability}

A third epoch of K-band photometry is available from the acquisition image
for the H+K-band spectroscopy in July 2002 (Appendix A2). The K-band magnitude
of LI-LMC 1813 is obtained by assuming that the K-band magnitude of KMHK
1603-1 remained unchanged at $K=11.80$ mag.

A second epoch of L$^\prime$-band photometry is obtained from the acquisition
image taken for the L-band spectroscopy in December 2001 (Appendix A2). The
L$^\prime$-band magnitude of KMHK 1603-1 is assumed to have remained unchanged
at $L^\prime=11.75$ mag (if the zero point were the same as that in October
2000, the L$^\prime$-band magnitude of KMHK 1603-1 would be
$L^\prime=11.80\pm0.30$ mag). Hence the L$^\prime$-band magnitude of LI-LMC
1813 is $L^\prime=9.27\pm0.25$ mag, in excellent agreement with the
flux-calibrated spectrum.

%
%
\begin{figure}
\psfig{figure=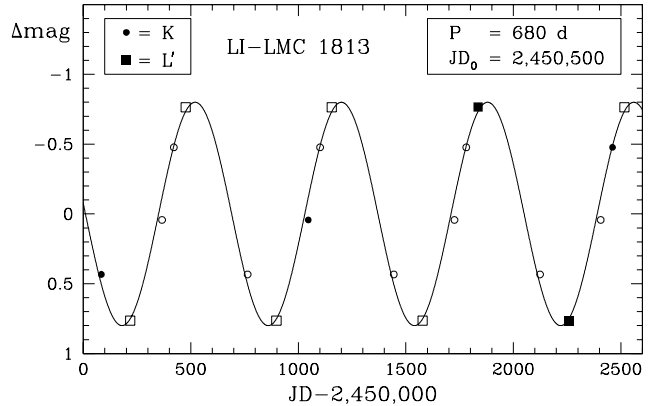,width=84mm}
\caption[]{K-band (solid disks) and L$^\prime$-band (solid squares) photometry
of LI-LMC 1813, with respect to their mean magnitudes, as a function of Julian
Date. The data can be matched by a sinusoidal curve with period $P=680$ d. The
open symbols are the data repeated each 680 d.}
\end{figure}

A sparse lightcurve is constructed from the differences of the three K-band
and two L$^\prime$-band magnitudes with their mean (Fig.\ 3): the variability
of (obscured) AGB stars in the K and L$^\prime$-bands are in phase with each
other, and have similar amplitudes (see Whitelock et al.\ 2003). The data are
consistent with a pulsation period of $P\sim$\,680 d. If confirmed, this would
be rather long but not unusual for highly evolved AGB stars (Wood 1998). The
large L$^\prime$-band amplitude of ${\Delta}L^\prime\sim$\,1.6 mag might be
due in part to temporal variations in the amount and temperature of the warm
dust contributing to the IR excess at $\lambda=$3--4 $\mu$m.

\subsection{Interstellar extinction}

The interstellar extinction towards the LMC is generally low (but see van Loon
et al.\ 1997 for severely reddened background galaxies). From the reddening
maps provided by Burstein \& Heiles (1982), based on galaxy counts, the
reddening towards LI-LMC 1813 is estimated to be $E(B-V)=0.06$ mag. The
reddening maps derived from the far-IR dust emission mapped by COBE/DIRBE
(Schlegel, Finkbeiner \& Davis 1998) yield a very consistent value of
$E(B-V)=0.056$ mag, which we adopt here. The spatial resolution of these maps
is insufficient to detect differences in reddening over a few arcminutes, and
therefore uniform reddening is assumed for the region within a radius of
$\sim2^\prime$ around KMHK 1603. This converts to an extinction of
$A_\lambda=0.20$, 0.15 and 0.11 in the g, r and i-bands, and $A_\lambda=0.05$,
0.03, 0.02 and 0.01 mag in the J, H, K and L$^\prime$-bands, respectively
(Mathis 1990). The (uncorrected) photometry of Table 1 is corrected for this
extinction before further analysis.

\subsection{The distance to LI-LMC 1813}

For the distance modulus of the LMC we adopt $(m-M)_0=18.55\pm0.17$ mag as
derived by Groenewegen (2000) from the Cepheid period-luminosity relationship.
However, the LMC is inclined, causing the distance modulus to vary across the
face of the LMC. This is especially important for LI-LMC 1813, which is
situated near the rim of the galaxy. Using the values for the centre,
inclination and position angle of the nodes from van der Marel \& Cioni
(2001), the differential distance modulus with respect to the centre of the
LMC is $\Delta(m-M)=-0.08$ mag. Within an uncertainty of $\sim$\,0.2 mag, the
distance modulus to LI-LMC 1813 then becomes $(m-M)=18.47$ mag.

\section{The progenitor mass of LI-LMC 1813}

The progenitor mass of LI-LMC 1813 may be estimated by isochrone-fitting to
the colour-magnitude diagram of the stars in the cluster KMHK 1603.

\subsection{Multi-object photometry}

Multi-object photometry was performed on the optical images using an
implementation of DAOPHOT (version {\sc ii}) and ALLSTAR (Stetson 1987) within
ESO-MIDAS. The reader is referred to van Loon et al.\ (2001) for a detailed
description of the Point Spread Function (PSF), multi-object photometry and
photometric calibration. As a result, $\sim7\times10^3$ sources were detected
in each of the gri-bands in the $3.7\times10^{-3}$ square degree area around
KMHK 1603. The completeness drops rapidly beyond $\sim$\,23 mag.

The point-sources were cross-correlated between the different filters after
determining and applying a geometric transformation (rotation and linear
translation), using an iterative scheme with a growing search radius and
rejecting extended sources on the basis of the sharpness parameter returned by
ALLSTAR. The photometry was then corrected for interstellar extinction as
derived in section 2.4.

Because there are only $\sim10^2$ objects visible in the near-IR images, and
hence these images are not at all crowded, the near-IR photometry was obtained
manually by aperture photometry. The photometry was transformed onto the SAAO
system (see Appendix A1.2) and then corrected for interstellar extinction
(section 2.4).

%
%
\begin{figure}
\psfig{figure=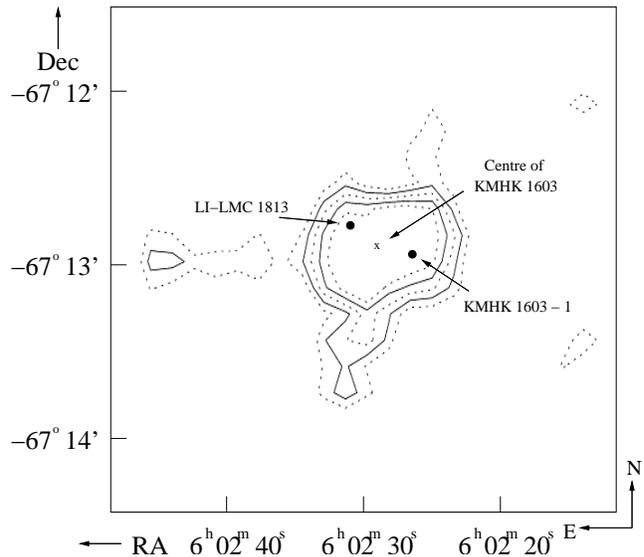,width=84mm}
\caption[]{Stellar surface density in the g-band. Contour levels range from
0.04 to 0.06 stars per square arcsecond, with steps of 0.005 stars per square
arcsecond. The centre of the cluster KMHK 1603, KMHK 1603-1 and LI-LMC 1813
are indicated.}
\end{figure}

The surface density of stars in the g-band image was measured by counting
stars with $g<22$ mag and sharpness $s \in [-0.5,0.5]$, within a
$28^{\prime\prime}\times28^{\prime\prime}$ box which was successively moved
across the image at steps of $\sim9^{\prime\prime}$. The resulting contour map
(Fig.\ 4) clearly shows the cluster KMHK 1603 as a region with a higher
density of stars as compared to the surrounding field. The cluster appears to
be somewhat elliptical, but for practical purposes we assume it to be
spherical. The centre of the cluster is determined to be located at (J2000)
RA$=6^{\rm h}02^{\rm m}29^{\rm s}$ and
Dec$=-67^\circ12^\prime54^{\prime\prime}$.

%
%
\begin{figure}
\psfig{figure=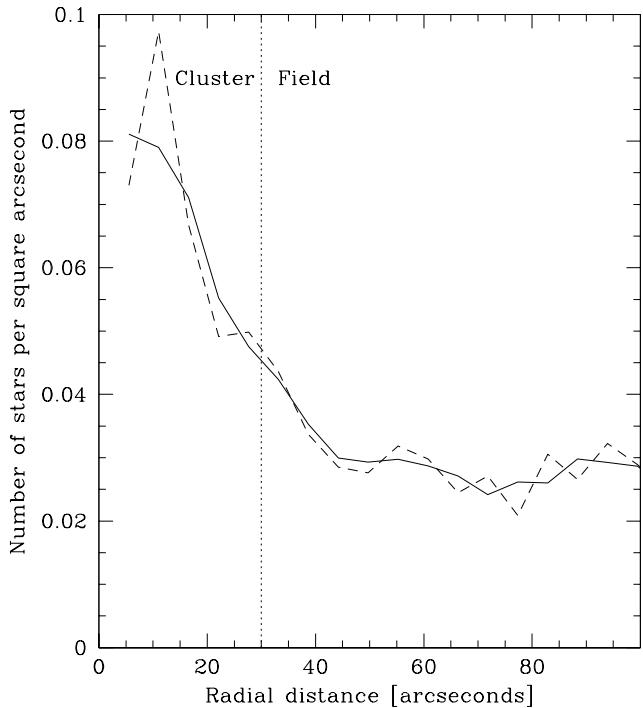,width=84mm}
\caption[]{Radial stellar surface density profile in the g-band (solid:
smoothed; dashed: unsmoothed), with respect to the centre of the cluster
KMHK 1603.}
\end{figure}

The surface density of stars, as measured in $6^{\prime\prime}$ annuli centred
on the cluster, is given as a function of the radial distance in Fig.\ 5. The
surface density in the core of the cluster is three times that of the field,
and it drops to the field level at a radius of $40^{\prime\prime}$. With a
(projected) distance to the cluster centre of $15^{\prime\prime}$, LI-LMC 1813
is located well within the cluster boundaries. To minimise contamination by
field stars, the colour-magnitude diagrams of the cluster were constructed for
stars within a smaller radius of $30^{\prime\prime}$.

\subsection{The age and metallicity of KMHK 1603}

%
%
\begin{figure*}
\psfig{figure=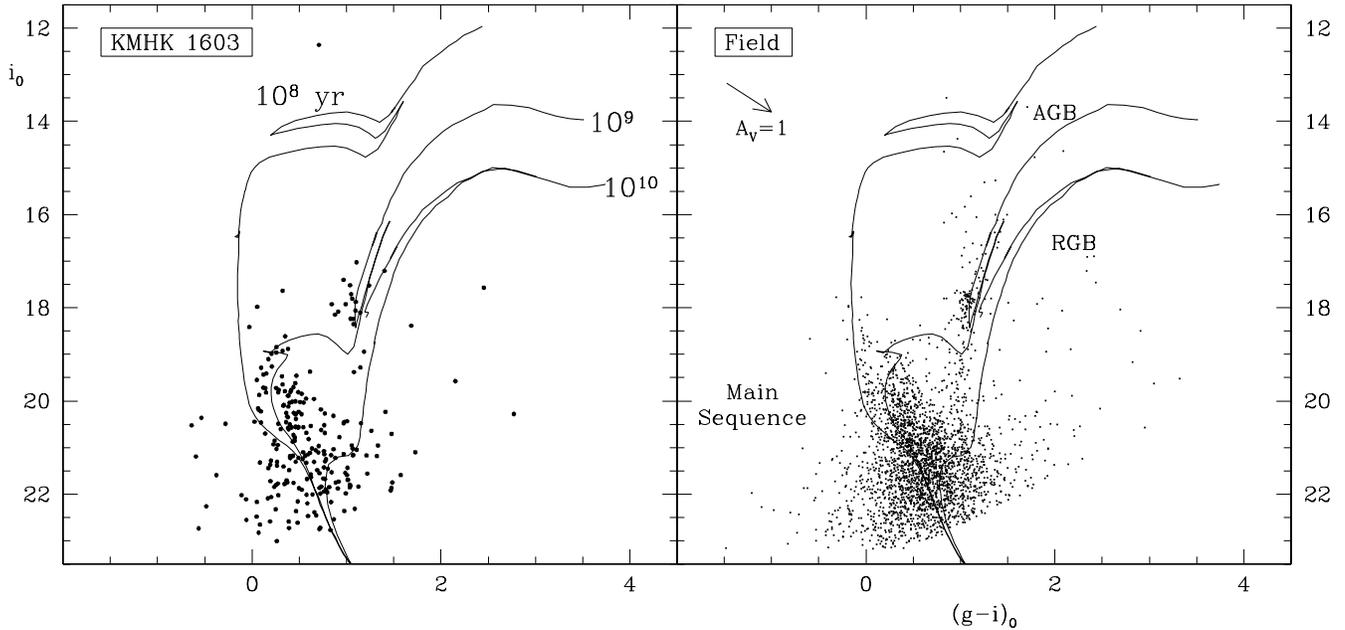,width=176mm}
\caption[]{Colour-magnitude diagram of $i_0$ versus $(g-i)_0$ for the LMC
cluster KMHK 1603 (left) and surrounding field (right). Isochrones (Bertelli
et al.\ 1994) are plotted for [M/H]$=-0.4$, and ages of $10^8$, $10^9$ and
$10^{10}$ yr. LI-LMC 1813 is not detected at these wavelengths.}
\end{figure*}

The interstellar extinction-corrected $i_0$ versus $(g-i)_0$ colour-magnitude
diagram is presented for the cluster KMHK 1603 (Fig.\ 6, left) and the
surrounding field (Fig.\ 6, right). Also plotted are three isochrones from
Bertelli et al.\ (1994), for a typical LMC metallicity of [M/H]$=-0.4$ and
ages of 10 Gyr, 1 Gyr and 100 Myr, after transformation onto the photometric
systems as described in detail in van Loon et al.\ (2003). These isochrones do
not include the effects of circumstellar extinction.

Both the cluster and field show a well-populated Main Sequence up to an age of
$t\sim10^9$ yr, and a red clump of core He-burning giants, at $(g-i)_0\sim$\,1
and $i_0\sim$\,18 mag, consistent with such intermediate or older ages.
Although there are a number of field Main Sequence stars brighter than the
$10^9$ yr Main Sequence turnoff, and corresponding RGB or AGB stars with
$i_0\sim$\,16 mag, there is a clear absence of stars younger than a few $10^8$
yr.

%
%
\begin{figure}
\psfig{figure=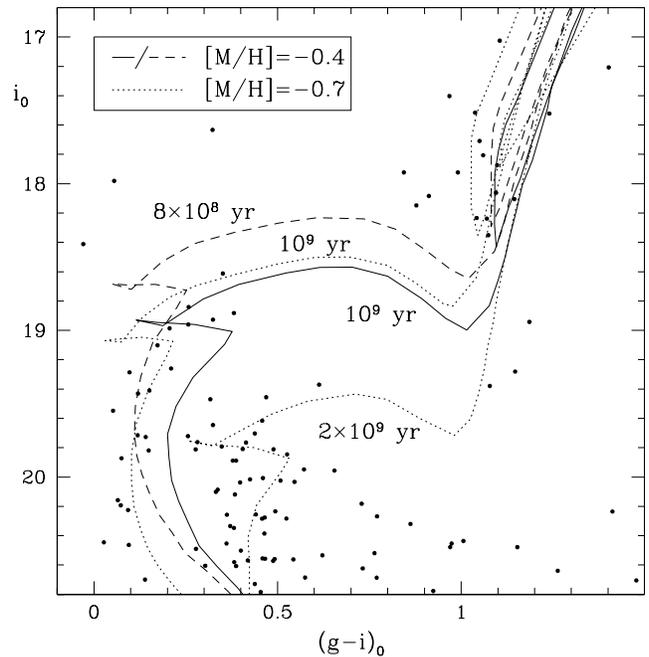,width=84mm}
\caption[]{Close-up of the Main-Sequence turnoff and red clump regions of the
colour-magnitude diagram of $i_0$ versus $(g-i)_0$ for the LMC cluster KMHK
1603. Isochrones (Bertelli et al.\ 1994) are plotted for several combinations
of age and metallicity.}
\end{figure}

A close-up of the optical colour-magnitude diagram for the cluster stars
(Fig.\ 7) reveals two populations: one of $t\sim$\,1 Gyr and an older one of
$t\sim$\,2 Gyr. The younger population is traced by a Main Sequence turnoff
(MS-TO) at $(g-i)_0\sim$\,0.2 and $i_0\sim$\,19.0 mag, clearly separated from
the older population with a MS-TO at $(g-i)_0\sim$\,0.4 and $i_0\sim$\,19.8
mag. Both populations show a red clump at roughly the same location in the
colour-magnitude diagram. The younger population stands out prominently
amongst the stars in the direction of the cluster, but much less so in the
field: counting stars with $(g-i)_0 \in [0.1,0.4]$ and $i_0 \in [18.5,19.3]$
mag ($\sim$\,1 Gyr MS-TO: 1) and stars with $(g-i)_0 \in [0.3,0.6]$ and $i_0
\in [19.4,20.2]$ mag ($\sim$\,2 Gyr MS-TO: 2), the ratio $N_1/N_2=0.21$ for
the field but 0.42 in the direction of the cluster. Hence we identify the
$t\sim$\,1 Gyr stars with the cluster population, and the $t\sim$\,2 Gyr stars
with the field population. This is in good agreement with the star formation
epochs around 1 and 2 Gyr that Smecker-Hane et al.\ (2002, their Fig.\ 5) find
for another field in the LMC disk. There is also evidence for a younger
population of several $10^8$ yr, both around KMHK 1603 (Fig.\ 6, right) and in
the disk field studied by Smecker-Hane et al.\ (2002).

A better match to the optical colour-magnitude diagram of the cluster stars is
obtained by an isochrone of slightly younger age, $t\sim$\,800 Myr, yet the
best match is obtained by an isochrone with $t\sim$\,1 Gyr but with a somewhat
lower metallicity of [M/H]$\sim-0.7$. Smith et al.\ (2002) also argue that,
with [M/H]$\sim-0.6$, LMC red giants with $M_{\rm ZAMS}\sim$\,3 M$_\odot$
often have metallicities lower than what is normally taken as the canonical
LMC metallicity, [M/H]$\sim-0.4$ to $-0.3$. Consequently, we also suspect that
the older, a few Gyr-old, field population has a rather low metallicity of
[M/H]$\sim-0.7$.

%
%
\begin{figure}
\psfig{figure=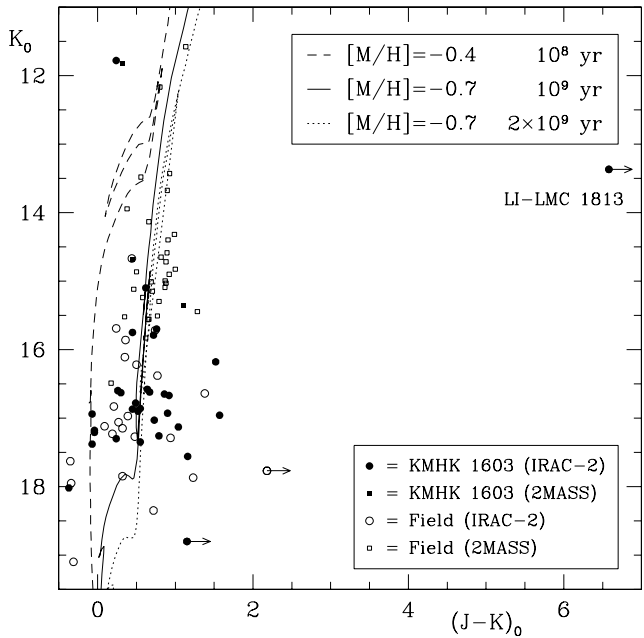,width=84mm}
\caption[]{Colour-magnitude diagram of $K_0$ versus $(J-K)_0$ for the LMC
cluster KMHK 1603 (solid) and surrounding field (open), both from the IRAC-2
imaging as well as from the 2MASS database. Isochrones (Bertelli et al.\ 1994)
are plotted for several combinations of age and metallicity. The AGB star
LI-LMC 1813 is extremely red due to circumstellar extinction.}
\end{figure}

The near-IR colour-magnitude diagram does not distinguish very well between
different intermediate-age populations (Fig.\ 8), but it is clear that the
cluster population is not consistent with an age of much less than $t\sim$\,1
Gyr (note that the 2MASS data sample a much larger area than the IRAC-2 data).
The red clump is clearly visible in the cluster, at $(J-K)_0\sim$\,0.5 and
$K_0\sim$\,17 mag.

The brightness of the red clump stars can be used as an age indicator. In the
optical, the red clump in the cluster KMHK 1603 is observed at $m_{\rm
i}=18.10\pm0.20$ mag (Fig.\ 7) or, assuming that $m_{\rm i}{\sim}m_{\rm I}$,
at $M_{\rm I}=-0.37\pm0.20$ mag. This suggests an age of either
$t=0.82^{+0.21}_{-0.14}$ Gyr or $t>$1.22 Gyr for [M/H]$=-0.4$, and
$t=0.96^{+\infty}_{-0.16}$ Gyr for [M/H]$=-0.7$ (Girardi \& Salaris 2001,
their Table 1). In the near-IR, the red clump is observed at $m_{\rm
K}=16.90\pm0.20$ mag, or $M_{\rm K}=-1.57\pm0.20$ mag. This suggests an age of
either $t=0.72^{+0.16}_{-0.12}$ Gyr or $t>$1.32 Gyr for [M/H]$=-0.4$, and
either $t=0.78^{+0.16}_{-0.12}$ Gyr or $t>$1.39 Gyr for [M/H]$=-0.7$ (Salaris
\& Girardi 2002, their Table 1). The ambiguity in the age solution is the
reason for the near-coincidence of the red clump for the cluster and field
stars, if the cluster and field are dominated by populations of $t\sim$\,1 Gyr
and $t>1.4$ Gyr, respectively.

\subsection{Conclusion: progenitor mass of LI-LMC 1813}

We conclude that LI-LMC 1813 probably has an age of $t=1$ Gyr (or slightly
younger) and an initial metallicity of [M/H]$=-0.7$ (or slightly higher), and
hence a progenitor mass of $M_{\rm ZAMS}=2.2\pm0.1$ M$_\odot$ (Bertelli et
al.\ 1994).


\section{Spectroscopic evidence for the carbon-rich nature of LI-LMC 1813}

Dusty circumstellar envelopes may be separated into carbon-rich and
oxygen-rich based on their location in IR colour-magnitude diagrams, but this
diagnostic can be ambiguous (van Loon et al.\ 1998; Trams et al.\ 1999b): the
colours of LI-LMC 1813, $(K-L)\sim$\,4.3 mag,
$(K-[12])=K+2.5\log{(F_{12}/28.3)}\sim$\,8 mag, and
$([12]-[25])=2.5\log{\left((F_{25}/6.73)\times(28.3/F_{12})\right)}\sim$\,2
mag, are consistent with both carbon and oxygen-rich circumstellar material
(Trams et al.\ 1999b, their Figs.\ 7 \& 8). To determine the dominant
chemistry of the stellar atmosphere, we thus resorted to taking near-IR
spectra in the H, K and L-bands. These observations are described in full in
Appendix A2.

%
%
\begin{figure}
\psfig{figure=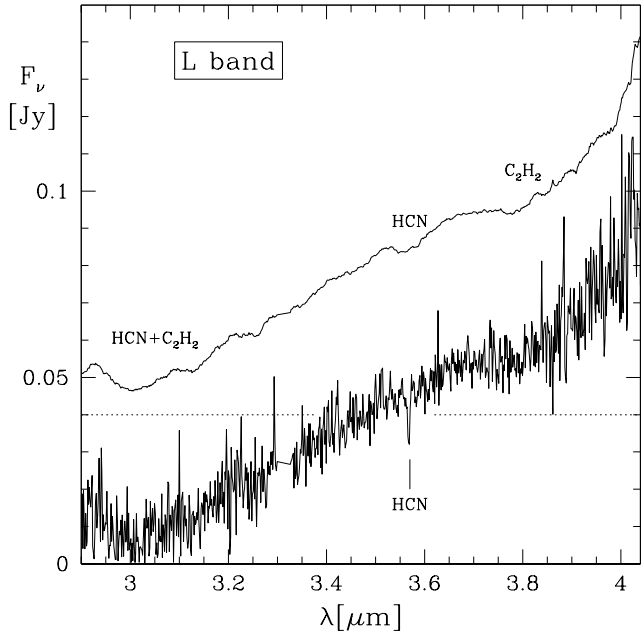,width=84mm}
\caption[]{L-band spectrum of LI-LMC 1813. A smooth version is offset for
clarity. Most of the narrow spikes in the unsmoothed spectrum are residuals
from the telluric line removal.}
\end{figure}

The L-band spectrum of LI-LMC 1813 (Fig.\ 9) features strong absorption around
3.1 $\mu$m from HCN and C$_2$H$_2$ molecules, a narrow absorption feature at
3.55 $\mu$m due to HCN, and absorption around 3.8 $\mu$m which is attributed
to C$_2$H$_2$ (Matsuura et al.\ 2002). These molecules only form in great
numbers after the supply of oxygen has been exhausted in the formation of CO.
Hence the strong absorption due to carbonaceous molecules is only seen in the
spectra of carbon stars, that have $n({\rm C})/n({\rm O})>1$. The strength of
the 3.55 and 3.8 $\mu$m absorption bands is remarkable, especially as they are
veiled by continuum emission from the circumstellar dust at wavelengths
$\lambda>3$ $\mu$m (see Fig.\ 12). This hints at a contribution from molecules
within the dust envelope surrounding the star, causing absorption seen against
the thermal dust continuum emission.

%
%
\begin{figure}
\psfig{figure=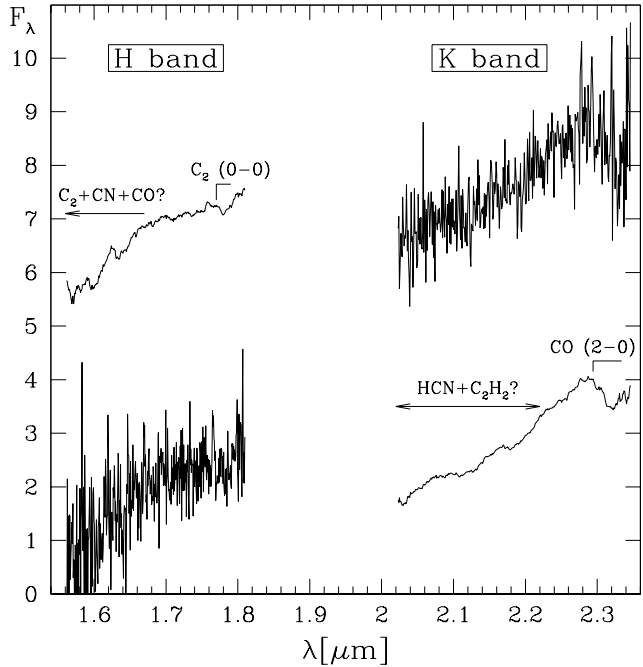,width=84mm}
\caption[]{H and K-band spectra of LI-LMC 1813, on an arbitrary flux density
scale. Smoothed versions of these noisy data are plotted offset, and some of
the most conspicuous absorption features are labelled.}
\end{figure}

The carbon-rich nature of LI-LMC 1813 is confirmed by the H and K-band spectra
(Fig.\ 10), which clearly show the absorption bandhead of CO (${\Delta}v=-2$)
at 2.293 $\mu$m, and the Ballik-Ramsay bandhead of C$_2$ (${\Delta}v=0$) at
1.77 $\mu$m. The strength of the latter requires a high $n({\rm C})/n({\rm
O})$ ratio, possibly $>1.4$ (Lan\c{c}on \& Wood 2000). A high ratio of $n({\rm
C})/n({\rm O})>1.2$ has also been invoked by Matsuura et al.\ (2002) to
explain the high abundance of C$_2$H$_2$ in the spectra of LMC carbon stars
--- including LI-LMC 1813. However, the molecular abundances not only depend
on the free carbon abundance, but also on the local gas temperature (Matsuura
et al., in preparation).

%
%
\begin{figure}
\psfig{figure=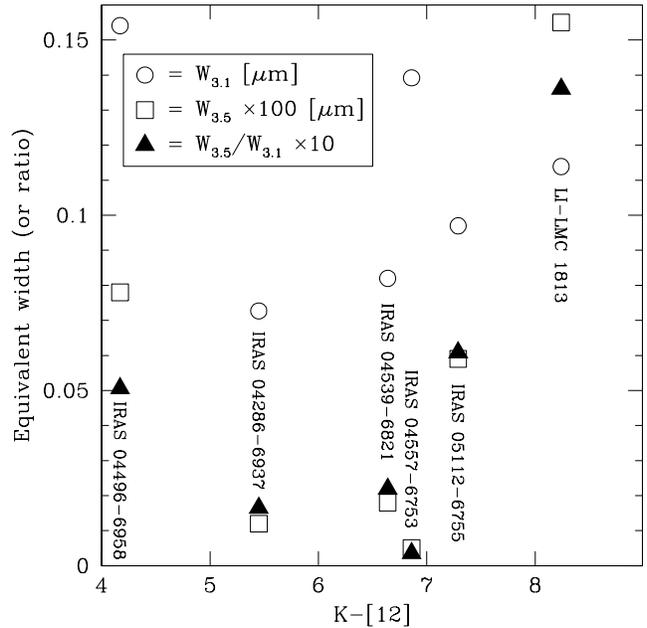,width=84mm}
\caption[]{Equivalent widths of the absorption due to HCN+C$_2$H$_2$ at 3.1
$\mu$m, $W_{3.1}$ (circles), and due to HCN at 3.55 $\mu$m, $W_{3.5}$
(squares), and their ratio, $W_{3.5}/W_{3.1}$ (triangles), for the LMC carbon
stars from Matsuura et al.\ (2002).}
\end{figure}

Although the equivalent width of the 3.1 $\mu$m absorption in LI-LMC 1813 is
very similar to that in other dust-enshrouded carbon stars in the LMC (Fig.\
11 and Matsuura et al.\ 2002), the equivalent width of its 3.55 $\mu$m
absorption is the largest of that sample. It also has the reddest $(K-[12])$
colour amongst these stars, which indicates a high mass-loss rate. These
observations suggest either a highly advanced evolutionary state accompanied
by a large $n({\rm C})/n({\rm O})$ ratio, or a very cool extended atmosphere
leading to the abundant formation of molecules.


\section{The mass-loss rate of LI-LMC 1813}

\subsection{Modelling the Spectral Energy Distribution}

%
%
\begin{figure}
\psfig{figure=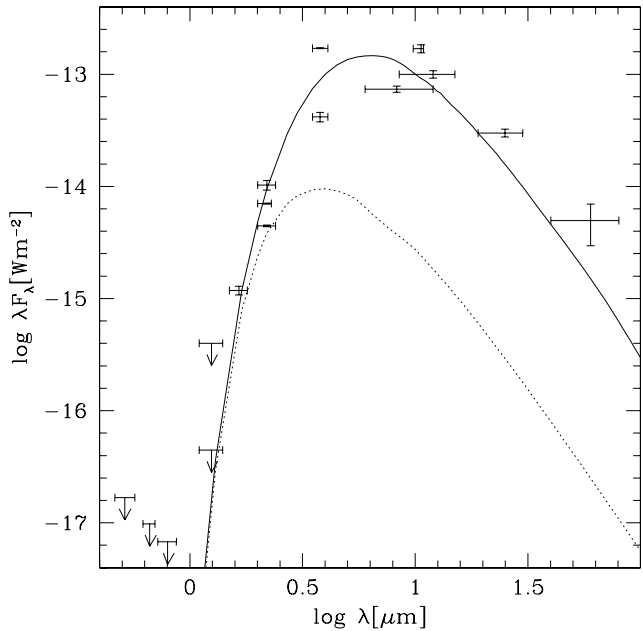,width=84mm}
\caption[]{Interstellar-extinction corrected spectrophotometric energy
distribution of LI-LMC 1813, and a Dusty model fit with $L=1.5\times10^4$
L$_\odot$ and $\tau_{\rm V}=50$ (solid). The stellar contribution after
circumstellar extinction is plotted too (dotted).}
\end{figure}

The SED was modelled with the radiative transfer code "Dusty" (Ivezi\'{c},
Nenkova \& Elitzur 1999). The input parameter "density type" was set to 3,
which gives a density structure consistent with radiation-driven wind theory.
The result of the best fit to the data is shown in Fig.\ 12.

A stellar temperature of $T_{\rm eff}=2500$ K is used as such cool photosphere
is typical for dust-enshrouded AGB stars (van Loon et al.\ 1998; Groenewegen
\& Blommaert 1998): carbon stars are produced with $T_{\rm eff}\sim$\,2900 K,
the superwind phase starts at $T_{\rm eff}\sim$\,2700 K, and the coldest
carbon stars known have $T_{\rm eff}\sim$\,2200 K (Marigo 2002). Such range in
temperatures corresponds to a range in derived luminosities of only $\pm$ 5\%,
and a range in mass-loss rates of $\pm$ 8\%.

The temperature at the inner radius of the dust shell is assumed to be $T_{\rm
inner}=1000$ K as this is the typical temperature at which dust grains
condense in a circumstellar environment. Since the data used to constrain the
SED were taken at many epochs, it is quite possible that variations in the
physical characteristics of the dust shell have occurred and therefore
modelling with one set of properties will only provide an accurate fit for a
portion of the SED. Stellar pulsations lead to variability in the luminosity,
and therefore the dust temperature. A closer fit to the IRAS 25 and 60 $\mu$m
data is obtained by assuming $T_{\rm inner}=800$ K. This would also increase
the stellar contribution in the L$^\prime$-band, where photospheric spectral
features are seen. Evidence of a higher dust temperature is provided by the
much brighter L$^\prime$-band photometric point. Averaged over the pulsational
period, a value of $T_{\rm inner}=1000$ K seems appropriate.

As LI-LMC 1813 is a carbon star (Section 4) we use amorphous carbon grains
(Zubko et al.\ 1996) ranging in size from $a=0.01$ to 0.1 $\mu$m according to
a power-law distribution of the form $n({\rm a}){\propto}a^{-q}$ with $q=3.5$
(Mathis, Rumpl \& Nordsieck 1977). We assume the Dusty default for the dust
grain specific mass density of $\rho_{\rm grain}=3$ g cm$^{-3}$. The Dusty
model assumes a gas-to-dust mass-ratio of $\xi=\rho_{\rm gas}/\rho_{\rm
dust}=200$. However, intermediate-age stars in the LMC may have a higher
ratio, as their metallicity can be several times lower than solar. This only
becomes important when calculating the mass-loss rate from the optical depth
and luminosity. In particular, the radial density distribution throughout the
radiatively-driven circumstellar outflow depends on the optical depth but not
on the gas-to-dust ratio (Elitzur \& Ivezi\'{c} 2001).

The shell thickness depends on the outflow velocity and the length of time for
which outflow has occured. The difference between the model results adopting
$r_{\rm outer}/r_{\rm inner}=10^2$ and $10^4$ is slight and only noticeable at
$\lambda>30$ $\mu$m. As there is no reason to assume that the dust shell is
either extremely thick or thin, we use a relative thickness of $10^3$. The
inner radius of the dust shell around LI-LMC 1813 is found to be $r_{\rm
inner}=3.9\times10^{12}$ m, and thus the outer radius lies at $r_{\rm
outer}=3.9\times10^{15}$ m. The Dusty model for the best fit to the SED of
LI-LMC 1813 predicts an outflow velocity of $v=9.5$ km s$^{-1}$, and hence the
duration of intense mass loss is estimated at $t\sim10^4$ yr. Although the
data do not constrain this timescale better than by a lower limit of roughly
$t>10^3$ yr, it does suggest that LI-LMC 1813 is undergoing a phase of
prolonged intense mass loss (superwind).

Dusty produces the shape of the emergent spectrum, which is then scaled to
match the observed flux. From the scaling factor we obtain the bolometric
luminosity of $L=1.5\times10^4$ L$_\odot$ for LI-LMC 1813, or $M_{\rm
bol}=-5.72$ mag. This agrees extremely well both with the original estimate of
$M_{\rm bol}=-5.78$ mag (van Loon et al.\ 1997), and with the value of $M_{\rm
bol, tip-AGB}=-5.75\pm0.04$ mag as predicted from the isochrone fit to KMHK
1603 (section 3.3).

The optical depth, $\tau$, not only determines the shape of the SED, but also
the mass-loss rate, $\dot{M}$, with $\dot{M}\propto\tau$. The optical depth at
visual wavelengths is found to be $\tau_{\rm V}=50$, reflecting the highly
obscured nature of LI-LMC 1813 (compare with van Loon et al.\ 1999b). The
Dusty model returns a value for the mass-loss rate valid for $L=10^4$
L$_\odot$ and $\xi=200$, with $\dot{M}$ depending on $L$ and $\xi$ as
$\dot{M}{\propto}L^{3/4}(\xi\rho_{\rm grain})^{1/2}$. At solar metallicity,
dust-enshrouded objects have $\xi\sim$\,100. However, the metallicity of
LI-LMC 1813 is $\sim$\,3--5 times less than solar (Section 3). As the
gas-to-dust ratio scales inversely proportional to the metallicity (van Loon
2000), we expect $\xi\sim$\,300 to 500 in the dust shell of LI-LMC 1813. Hence
we estimate the mass-loss rate of LI-LMC 1813 to be
$\dot{M}=3.7(\pm1.2)\times10^{-5}$ M$_\odot$ yr$^{-1}$.

\subsection{Discussion: the superwind of the metal-poor carbon star LI-LMC
1813}

LI-LMC 1813 is a carbon star. With its estimated progenitor mass of $M_{\rm
ZAMS}=2.2$ M$_\odot$, $3^{\rm rd}$ dredge-up is indeed expected to have worked
(Marigo 2001) in the absence of HBB (van Loon et al.\ 2001) to raise the
photospheric abundance of carbon to above that of oxygen. The observed high
$n({\rm C})/n({\rm O})$ ratio can be explained in terms of its low metallicity
and relatively high mass as compared to carbon stars in the solar
neighbourhood that have generally low-mass progenitors of $M_{\rm ZAMS}<2$
M$_\odot$ (see Marigo 2002, her Fig.\ 6). It also suggests that LI-LMC 1813
has been a carbon star for already quite some time.

LI-LMC 1813 is indeed a highly evolved AGB star. The isochrones that yield the
mass estimate (based upon the colour-magnitude diagram of the cluster stars)
for LI-LMC 1813 also indicate that its luminosity places it at the very tip of
its evolution along the AGB. Its luminosity of $L=1.5\times10^4$ L$_\odot$ is
a few times brighter than the peak in the optically bright carbon star
luminosity function (Groenewegen 2002), but the same as the peak in the
dust-enshrouded carbon star luminosity function (van Loon et al.\ 1999b). This
is not unexpected, as the duration of the TP-AGB is longest for stars in the
range $M_{\rm ZAMS}\sim$\,2 to 2.5 M$_\odot$ (Marigo 2001).

It is therefore not surprising that LI-LMC 1813 is in the superwind phase,
experiencing mass loss at a tremendous rate of $\dot{M}=3.7\times10^{-5}$
M$_\odot$ yr$^{-1}$. Schr\"{o}der, Winters \& Sedlmayr (1999) show that
low-mass carbon stars, with $M_{\rm ZAMS}<1.3$ M$_\odot$, expell thick shells
during a brief period of enhanced luminosity as a result of a thermal pulse,
but that more massive carbon stars such as LI-LMC 1813 suffer a longer period
of a persistently high mass-loss rate. Our analysis of LI-LMC 1813 provides
direct evidence for this superwind to occur at the {\em tip} of the AGB.

Vassiliadis \& Wood (1993) model TP-AGB evolution including a prescription for
the mass-loss rate based on empirical data for galactic Mira variables. They
argue that the maximum mass-loss rate during the superwind phase is set by the
transfer of momentum from the stellar radiation field onto the outflowing
matter: $\dot{M}_{\rm max}=v_{\rm exp}L/c$, independent of the metallicity or
chemical composition. Their prediction for the mass-loss rate of
$\dot{M}=1.2$--$1.4\times10^{-5}$ M$_\odot$ yr$^{-1}$ during the superwind
experienced by stars with $M_{\rm ZAMS}=1.5$--2.5 M$_\odot$ is three times
lower than the actual observed mass-loss rate of LI-LMC 1813. However, Gail \&
Sedlmayr (1986) have argued that multiple scattering can enhance the transfer
of momentum and thus increase the maximum mass-loss rate, which was confirmed
by observations of dust-enshrouded AGB stars in the LMC (van Loon et al.\
1999b).

Wachter et al.\ (2002) derive a formula for the mass-loss rate of
solar-metallicity carbon stars, based on dynamical models:
$\log{\dot{M}}=8.86-1.95\log{M/{\rm M}_\odot}-6.81\log{T_{\rm eff}/{\rm
K}}+2.47\log{L/{\rm L}_\odot}$. According to this formula, the mass-loss rate
of LI-LMC 1813 is predicted to be $\dot{M}\sim2^{+4}_{-1}\times10^{-5}$
M$_\odot$ yr$^{-1}$ for $T_{\rm eff}=2500\pm300$ K. This agrees extremely well
with the observed mass-loss rate of $\dot{M}=3.7(\pm1.2)\times10^{-5}$
M$_\odot$ yr$^{-1}$. Considering that the metallicity of LI-LMC 1813 is three
to five times lower than solar, the good agreement between the actual
mass-loss rate of LI-LMC 1813 and the prediction for its mass-loss rate at
solar metallicity suggests that the mass-loss rate --- at least during the
superwind phase --- does not strongly depend on metallicity (see van Loon
2000).

In her new calculations for AGB evolution and the formation of carbon stars,
Marigo (2002) includes the effects of molecule formation within the stellar
atmosphere. One of the differences between these and previous models is that
the $T_{\rm eff}$ drops once molecules are being formed, leading to high
mass-loss rates of several $10^{-5}$ M$_\odot$ yr$^{-1}$ at $n({\rm C})/n({\rm
O})>1.3$ for carbon stars of $M_{\rm ZAMS}>2$ M$_\odot$. The observed
insensitivity of $\dot{M}$ to the initial metallicity may, in the case of
carbon stars, be due to enhanced formation of carbonaceous molecules as a
result of a higher $n({\rm C})/n({\rm O})$ ratio at lower metallicity.
Alternatively, metallicity insensitivity of the pulsation mechanism has been
suggested as a possible reason for the metallicity insensitivity of the
mass-loss rate during the superwind phase (van Loon 2002).

When an AGB star arrives on the Mira period-luminosity (P-L) sequence, its
mass-loss rate has become so severe that further evolution is along tracks of
constant luminosity, with the radius (and hence pulsation period) growing as
the structure of the mantle re-adjusts itself to the decrease in mass (e.g.\
Wood 1998). Describing the Mira as a polytrope, $R\sqrt[3]{M}=constant$, and
the pulsation as a harmonic oscillator, $2\pi/P \simeq 1/t_{\rm freefall}
\simeq g(R)/R$, the period evolves as $P{\propto}1/M$ (van Loon 2002).

If LI-LMC 1813 has evolved from $P\sim$\,400 d on the Mira P-L ($M_{\rm
ZAMS}=2.2$ M$_\odot$) to $P\sim$\,680 d at present, it has a current mass of
only $M=1.3$ M$_\odot$. According to Marigo (2001), its final mass will become
$M_{\rm WD}=0.6$ M$_\odot$. This means that, although it has already lost
${\Delta}M\sim$\,0.9 M$_\odot$, it must still lose another
${\Delta}M\sim$\,0.7 M$_\odot$. Stars of $M_{\rm ZAMS}=2.2$ M$_\odot$ with
correspondingly long periods ($P\sim$\,1500 d) have not (yet) been found, the
longest known period for such star being $P\sim$\,900 d (Wood 1998). This is
not surprising, as it would perhaps take as little as $10^4$ yr to lose the
remainder of the stellar mantle, rendering these objects extremely rare.

\section{Summary}

As a result of a large observational effort, the mid-IR point source LI-LMC
1813 is now one of the few AGB stars experiencing the superwind phase of
intense mass loss for which the progenitor mass, initial metallicity,
atmospheric chemistry, luminosity and mass-loss rate are accurately known. It
thus provides an important testcase against which to gauge models for the
evolution and mass-loss of AGB stars.

Absorption features due to carbonaceous molecules in the near-IR spectra of
LI-LMC 1813 indicate that it is a carbon star. From its location in the LMC
cluster KMHK 1603, its progenitor mass is inferred to be $M_{\rm ZAMS}=2.2$
M$_\odot$, which is typical for a carbon star, whilst its initial metallicity
is three to five times lower than solar.

The object is heavily obscured by its circumstellar dust, which shines
brightly at IR wavelengths. From the modelling of the spectral energy
distribution, a mass-loss rate of $\dot{M}=3.7\times10^{-5}$ M$_\odot$
yr$^{-1}$ is estimated. This is very high, especially for such metal-poor
star. As the luminosity of LI-LMC 1813 ($L=1.5\times10^4$ L$_\odot$) places it
at the very tip of its AGB, this observation provides direct evidence for the
superwind to represent the final episode in the AGB evolution.

Although the observed mass-loss rate for LI-LMC 1813 is a few times higher
than that predicted by the formalism in Vassiliadis \& Wood (1993), it is in
very good agreement with the prescription based on the dynamical models of
Wachter et al.\ (2002) for {\em solar} metallicity. This suggests that the
mass-loss rate in the superwind phase may be metallicity insensitive (van Loon
2000, 2002), perhaps because of a larger abundance of carbonaceous molecules
other than carbon-monoxide in metal-poor stellar atmospheres (Matsuura et al.\
2002; Marigo 2002).

\section*{Acknowledgments}
We are grateful to the former Director of ESO for the generous allocation of
Director's Discretionary Time on the 0.9m Dutch telescope from Christmas to
New Year's Eve 1996, to Maria Messineo for taking the H+K-band spectra of
LI-LMC 1813 at CTIO (and commenting on an earlier version of this manuscript),
and to Dr.\ George Hau and the ESO 3.6m team for taking the optical spectrum
of KMHK 1603-1 during an EFOSC-2 set-up night. We also would like to thank
Dr.\ Ren\'{e} Mendez (ESO) for digging up the ancient IRAC-2 filter
transmission curves. This publication makes use of data products from the Two
Micron All Sky Survey, which is a joint project of the University of
Massachusetts and the Infrared Processing and Analysis Center/California
Institute of Technology, funded by the National Aeronautics and Space
Administration and the National Science Foundation. This research also made
use of data products from the Midcourse Space Experiment. Processing of the
data was funded by the Ballistic Missile Defense Organization with additional
support from NASA Office of Space Science. This research has also made use of
the NASA/IPAC Infrared Science Archive, which is operated by the Jet
Propulsion Laboratory, California Institute of Technology, under contract with
the National Aeronautics and Space Administration. Jacco thanks, above all,
Joana... for everything.

\appendix

\section{Observations}

\subsection{Photometry}

\subsubsection{Optical imaging (Dutch 0.9m, La Silla)}

The direct imaging camera at the Dutch 0.9m telescope at La Silla, Chile, was
used on the six nights of December 25 to 30, 1996, to obtain deep images of a
$3.77^\prime\times3.77^\prime$ region around the cluster KMHK 1603, through
Gunn g ($\lambda_0=5148$ \AA, $\Delta\lambda=81$ \AA), r ($\lambda_0=6696$
\AA, $\Delta\lambda=103$ \AA) and i ($\lambda_0=7972$ \AA, $\Delta\lambda=141$
\AA) filters (Thuan \& Gunn 1976; Wade et al.\ 1979). The total integration
time amounts to 4 hours per filter, split into 5-minute exposures to avoid
saturation and to allow for refocussing in order to reach the best image
quality. The pixels measure $0.442^{\prime\prime}\times0.442^{\prime\prime}$
on the sky, and stellar images on the combined (shift-added) frames have a
Full Width at Half Maximum (FWHM) of $\sim1.5^{\prime\prime}$, though some
individual frames show stellar images of $\lsim1.1^{\prime\prime}$ FWHM. The
CCD frames were reduced using standard procedures within the ESO-MIDAS
package. The reader is referred to van Loon et al.\ (2001) for more details.

\subsubsection{J and K-band imaging (IRAC-2, La Silla)}

The near-IR camera IRAC-2b at the ESO/MPI 2.2m telescope at La Silla, Chile,
was used on January 2, 1996, to image the field around the IRAS point source
LI-LMC 1813. Lens C was chosen to cover a field of view of
$133^{\prime\prime}\times 133^{\prime\prime}$, with a pixel scale of
$0.49^{\prime\prime}$. A sequence of 12 images in the K-band filter were
obtained, each consisting of ten 3-second exposures, shifted by
$5^{\prime\prime}$ in Right Ascension (RA) with respect to the previous image.
This procedure was then repeated backwards, using the J-band filter. The image
is deepest in the central $\sim75^{\prime\prime}\times 130^{\prime\prime}$.
The photometry is transformed onto the SAAO system (Carter 1990). The reader
is referred to van Loon et al.\ (1997) for more details.

\subsubsection{Near-IR imaging (2MASS \& DENIS)}

Archival J, H and K-band data is available from the second incremental data
release of 2MASS. It contains 8 stars within $1^\prime$ of LI-LMC 1813 which
were detected in the K band down to $K_{\rm limit}\sim$\,15 mag, including
LI-LMC 1813 itself, on 26 October 1998. Especially valuable is the detection
of LI-LMC 1813 in the H band, which is thereby the shortest wavelength at
which the object has been seen. The Denis point source catalogue towards the
Magellanic Clouds (Cioni et al.\ 2000) provides additional I, J and K$_{\rm
s}$-band data, but its sensitivity only reaches $K_{\rm s, limit}\sim$\,12 mag
and no star was detected within $1^\prime$ of LI-LMC 1813 except for the
optically brightest star in the cluster, KMHK 1603-1, which was detected in
the I and J-bands. The photometry from both 2MASS and Denis is transformed
onto the SAAO system (Carter 1990; Carpenter 2001) before further analysis.

\subsubsection{L$^\prime$-band imaging (ISAAC, Paranal)}

The Infrared Spectrometer And Array Camera (ISAAC) at the ESO VLT at Paranal,
Chile, was used on October 19, 2000, to image LI-LMC 1813 through an
L$^\prime$-band filter ($\lambda_0=3.78$ $\mu$m, $\Delta\lambda=0.58$ $\mu$m).
Because of the high background, the image was obtained after chopping and
nodding (both by $10^{\prime\prime}$ in Declination), with 9 sub-integrations
of 0.104 second at each chop position, and a total on-source integration time
of 1 minute. The field-of-view of $72^{\prime\prime}\times72^{\prime\prime}$
(at $0.071^{\prime\prime}$ pixel$^{-1}$) ensured that KMHK 1603-1 was imaged
simultaneously with LI-LMC 1813. Conditions were very good, with 6\% relative
humidity at a temperature of $15^\circ$C. The seeing was $0.67^{\prime\prime}$
at Zenith, with the target at an airmass of 1.442.

The data were reduced using ESO-MIDAS and the Eclipse package (Devillard
1997). First the electrical ghosts were removed ({\sc is\_ghost} in Eclipse).
Then the relative pixel responses were calibrated using a flatfield image, and
aperture photometry was performed on the difference of the two images at each
nodding position. The photometry was calibrated against that of the G3-giant
standard star HR 2354 ($L^\prime=4.989$ mag; van der Bliek, Manfroid \&
Bouchet 1996: virtually identical filter as that of ISAAC) at similar airmass
(1.392) and only 10 minutes after LI-LMC 1813.

\subsubsection{Mid-IR imaging (MSX)}

LI-LMC 1813 is included in the Mid-course Space eXperiment (MSX) Point Source
Catalogue (Price et al.\ 2001). It was only detected in band A (effective
wavelength $\lambda_0=8.28$ $\mu$m). The transmission of this band runs from
$\lambda\sim$\,6.1 to 10.8 $\mu$m, though, peaking at $\lambda_{\rm
peak}\sim$\,9.9 $\mu$m, and the measured flux density thus depends heavily on
the spectral slope of the object --- which is greatly affected by
circumstellar emission.

\subsubsection{N-band imaging (TIMMI-2, La Silla)}

The TIMMI-2 instrument at the ESO 3.6m telescope at La Silla, Chile, was used
on January 19, 2001, to obtain an image of LI-LMC 1813 through the N2 filter
($\lambda_0=10.6$ $\mu$m, with a transmission $>50$\% between $\lambda=9.74$
and 11.33 $\mu$m). At a pixel scale of $0.3^{\prime\prime}$ pixel$^{-1}$, the
slightly vignetted field measures $96^{\prime\prime}$ (RA) $\times$
$72^{\prime\prime}$ (Dec). The standard technique for observing in the thermal
IR was employed, chopping $20^{\prime\prime}$ in Dec and nodding
$20^{\prime\prime}$ in RA. The total on-source integration time was 15
minutes. The images were combined, and the photometry was obtained, using
standard commands within ESO-MIDAS. The K3 giant $\alpha$ Hya was used as a
prime calibrator, adopting a flux density of $F_{\rm N2}=129$ Jy (Cohen 1998).
The photometry was checked against the IRAS 12 $\mu$m flux densities for the
K2 and K1 giants $\lambda_2$ Tuc ($F_{12}=3$ Jy) and $\gamma$ Pic ($F_{12}=7$
Jy).

\subsubsection{Mid-IR photometry (IRAS)}

Data at 12, 25 and 60 $\mu$m were retrieved from the IRAS data base server in
Groningen\footnote{The IRAS data base server of the Space Research
Organisation of the Netherlands (SRON) and the Dutch Expertise Centre for
Astronomical Data Processing is funded by the Netherlands Organisation for
Scientific Research (NWO). The IRAS data base server project was also partly
funded through the Air Force Office of Scientific Research, grants AFOSR
86-0140 and AFOSR 89-0320.} (Assendorp et al.\ 1995). The Gipsy data analysis
software was used to measure the flux density from a trace through the
position of the star (Gipsy command {\sc scanaid}). For a discussion about the
sensitivity and reliability of this procedure for point sources in the LMC,
see van Loon et al.\ (1999b).

\subsection{Spectroscopy}

\subsubsection{Optical spectroscopy (EFOSC-2, La Silla)}

The EFOSC-2 instrument at the ESO 3.6m telescope at La Silla, Chile, was used
on October 31, 2002, to obtain an optical spectrum of KMHK 1603-1 between
$\lambda=4700$ and 6700 \AA\ with grism \#9. The slit was narrowed to a width
of only $0.5^{\prime\prime}$ projected on the sky, in order to boost the
spectral resolving power to $R\sim$\,2000. Two spectra were taken, with
exposure times of 300 seconds each, which were then combined. The bias level
was subtracted first. Differences in illumination and responsivity across the
image were corrected using flatfield spectra of an internal quartz lamp. The
wavelength axis was calibrated using two exposures of a He/Ar lamp, with an
accuracy to well within the spectral resolution. The sky-subtracted spectra
were corrected for extinction by the Earth's atmosphere, and for the response
function as derived from a spectrum of the HST standard star HD 49798.

\subsubsection{H and K-band spectroscopy (OSIRIS, CTIO)}

The Ohio State InfraRed Imager/Spectrometer (OSIRIS) at the 4m Blanco
telescope at Cerro Tololo (CTIO), Chile, was used on 6 July, 2002, to obtain a
spectrum of LI-LMC 1813 covering the 0.9 to 2.4 $\mu$m wavelength range. The
instrument was used first in imaging mode to acquire the star in the slit. The
spectra were acquired in cross-dispersed mode through a
$1^{\prime\prime}\times30^{\prime\prime}$ slit, providing a resolving power of
$R=1200$. Four spectra were taken, with each time the star moved over
$5^{\prime\prime}$ along the slit. The slit orientation was East--West.

The data reduction was performed using Iraf. In the individual spectra bad
pixels were eliminated and flatfield and linearisation corrections were
applied. Since the spectra were taken in the morning almost at sunrise, the
sky brightness changed considerably during the course of the observations.
From each image another image (the nearest in time) was subtracted to perform
a first sky subtraction. The individual spectra were then extracted and
combined. Stars of spectral type A and B were observed as close to the LI-LMC
1813 spectrum's airmass (1.756) as possible, in order to correct for telluric
absorption features and the instrumental response. The raw spectra of these
reference stars were obtained in the same manner as LI-LMC 1813. The hydrogen
and helium series appearing in these spectra were removed by linear
interpolation. The cleaned spectra of the early-type stars were divided by a
blackbody curve to eliminate the slope due to their intrinsic energy
distributions (see Messineo et al., in preparation, for more details).
Temperatures were adopted from Gray (1992) and Vacca, Garmany \& Schull
(1996). The wavelength calibration was obtained using the telluric OH lines
(Iraf tasks {\sc refspectra} and {\sc dispcor}; see Oliva \& Origlia 1992).
The spectra were dereddened (Iraf task {\sc deredden}) using the extinction
law of Cardelli, Clayton \& Mathis (1989) --- essentially the same as that of
Mathis (1990).

\subsubsection{L-band spectroscopy (ISAAC, Paranal)}

ISAAC at the ESO VLT at Paranal, Chile, was used on December 14, 2001, to take
an L-band spectrum of LI-LMC 1813. The total on-source exposure time was 36
minutes. With a slit width of $0.6^{\prime\prime}$, the resolving power was
$R\sim$\,600. The optical seeing was $1.2^{\prime\prime}$, but the infrared
seeing was $\sim0.5^{\prime\prime}$. The background was subtracted by chopping
with a chopping throw of $15^{\prime\prime}$. Telescope nodding and jitter was
used to correct for bad pixels. The data was reduced using the Eclipse package
(Devillard 1997). Variations of the pixel response across the array were
corrected with a flatfield of the twilight sky. The wavelength calibration is
based on the exposure by an Ar+Xe arc lamp with the same wavelength setting as
the target observation.

The weather was photometric. The B9 IV star HIP 029729 was used as a telluric
standard and flux calibration star. For the flux calibration we use the
effective temperature, $T_{\rm eff}$, and $(K-L)$ colour for a spectral type
B9 V instead of B9 IV, because no accurate value for $T_{\rm eff}$ is
available for sub-giants. The spectrum is represented by a black body with
$T_{\rm eff}=10,700$ K (Tokunaga 2000). The K-band magnitude of this star is
$K=7.536$ mag (van der Bliek et al.\ 1996) and $(K-L)=-0.03$ mag for a B9 V
star, hence the L-band magnitude is estimated to be $L=7.566$ mag. The L-band
magnitude is converted into Jansky according to the zero-magnitude in van der
Bliek et al.\ (1996).

\section{The optically bright star KMHK 1603-1}

The optical and infrared colours and magnitudes of the bright star KMHK 1603-1
are consistent with those of an F-type supergiant in the LMC. It could then
either be a field supergiant of relatively young age ($t\ll10^8$ yr), or a
post-AGB star in the cluster KMHK 1603. The latter case would be especially
interesting, as we could then compare the properties of an AGB and a post-AGB
star of the same main-sequence mass and metallicity. However, it could also be
a galactic foreground main-sequence star.

%
%
\begin{figure}
\psfig{figure=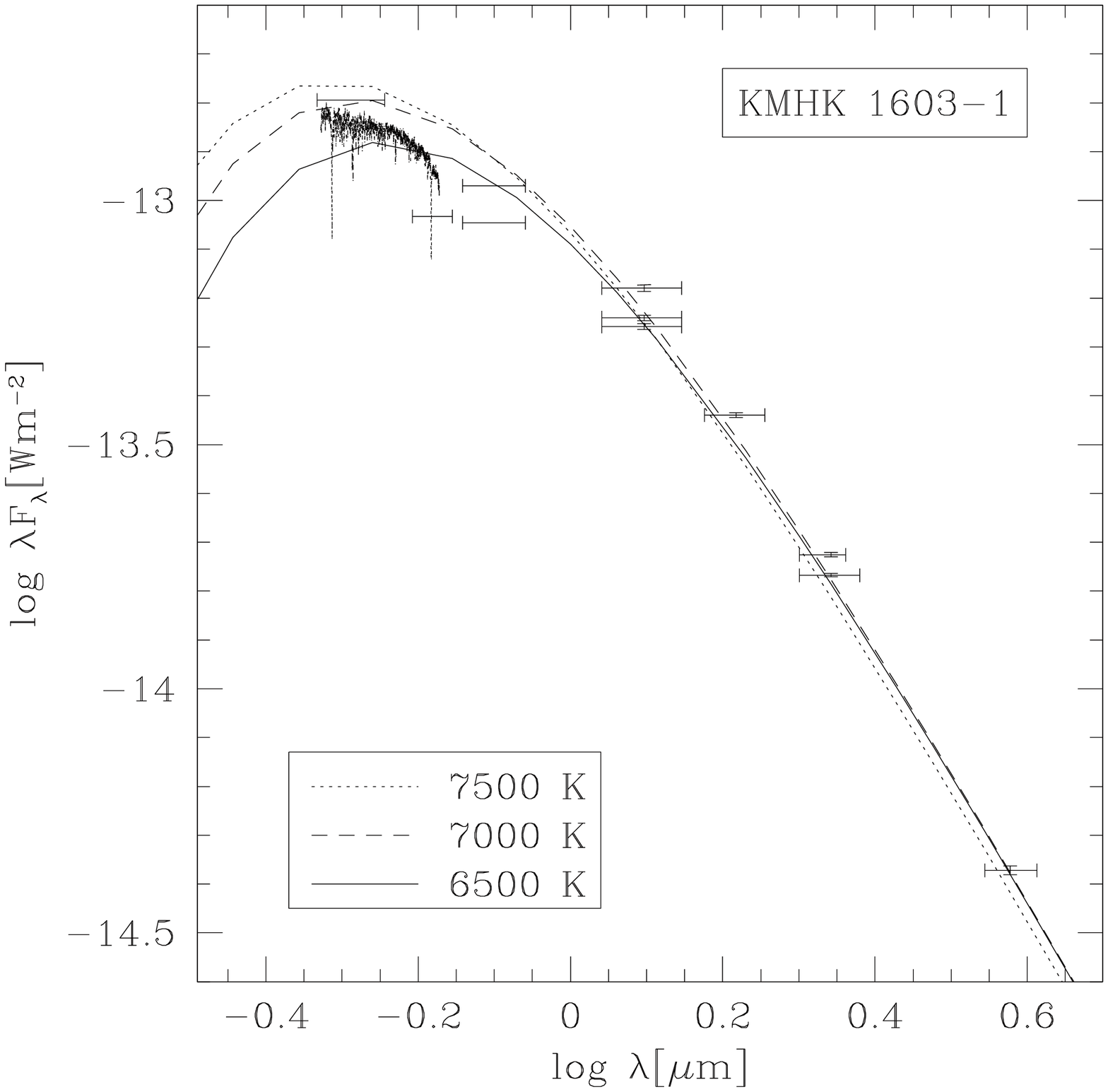,width=84mm}
\caption[]{Spectro-photometric energy distribution of KMHK 1603-1, corrected
for interstellar extinction, assuming that it is located in the LMC.
Blackbodies of $T=6500$, 7000 and 7500 K are shown for comparison.}
\end{figure}

Modelling of the photometry (Fig.\ B1) suggested a temperature of (roughly)
$T_{\rm eff}\sim$\,6500--7000 K. If at the distance of the LMC, it will then
have a bolometric luminosity of $L\sim1.5\times10^4$ L$_\odot$, i.e.\
virtually identical to that of LI-LMC 1813. As the luminosity of a post-AGB
star (before reaching the White Dwarf cooling track) is the same as it had
when it left the AGB, this lended further support for the hypothesis that KMHK
1603-1 is a post-AGB star in the KMHK 1603 cluster. Therefore, an optical
spectrum was taken in order to confirm or reject this.

%
%
\begin{figure}
\psfig{figure=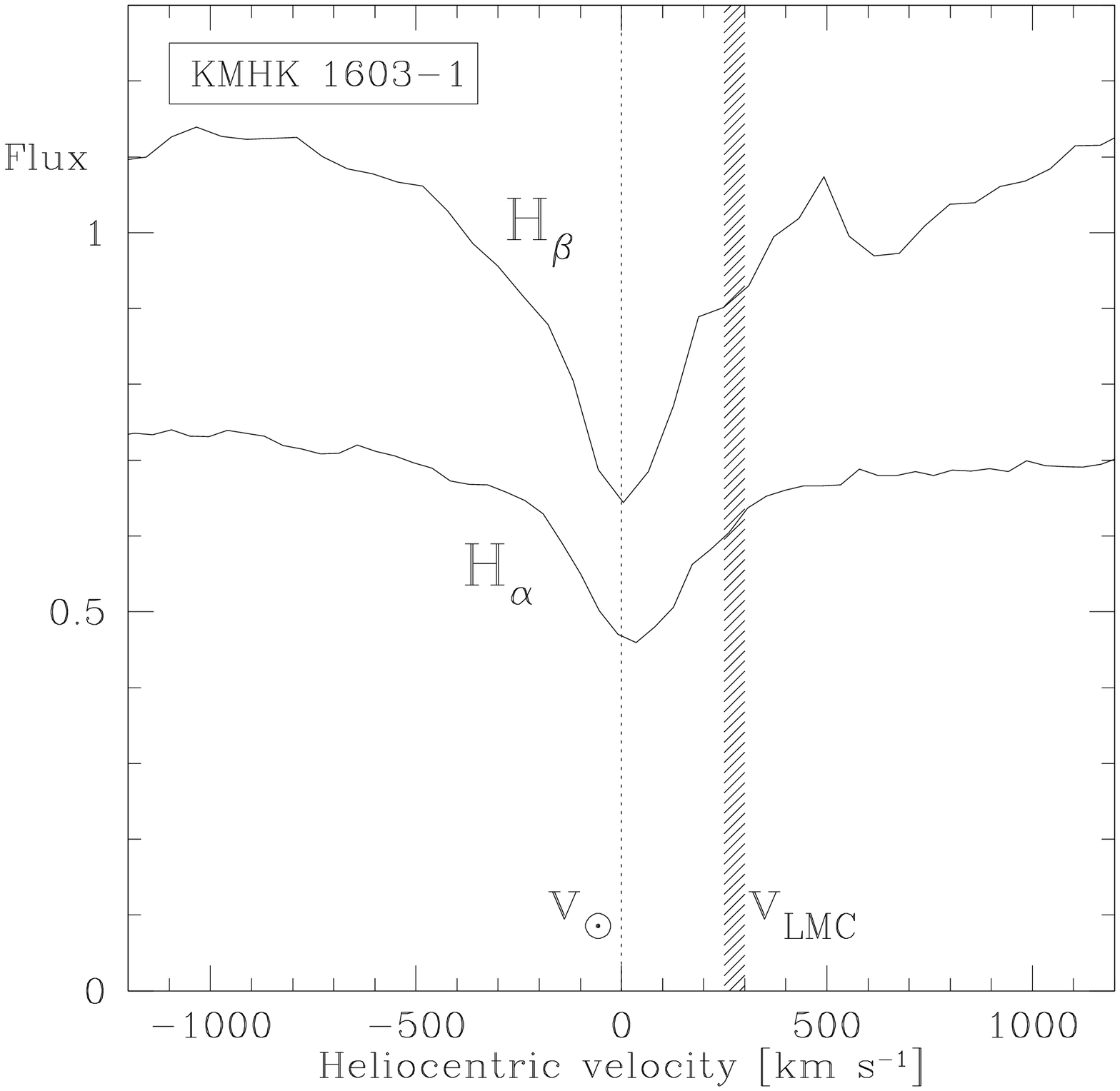,width=84mm}
\caption[]{H$\alpha$ and H$\beta$ line profiles in the optical spectrum of
KMHK 1603-1. Peaking to within $\sim$\,30 km s$^{-1}$ from the heliocentric
rest velocity, they rule out LMC membership.}
\end{figure}

The H$\alpha$ and H$\beta$ photospheric absorption lines in the spectrum of
KMHK 1603-1 peak within $\sim$\,30 km s$^{-1}$ from the heliocentric rest
velocity (Fig.\ B2), ruling out LMC membership. On the basis of the relative
strengths of the H$\alpha$ and H$\beta$ lines, $\lambda5174$ Mg~{\sc i}
triplet (Pritchet \& van den Bergh 1977), $\lambda\lambda5890,5896$ Na~{\sc i}
D doublet and the $\lambda5270$ and $\lambda6495$ Fe~{\sc i} lines, and by
comparison with the spectral library of Jacoby, Hunter \& Christian (1984),
the spectral type is estimated to be F6 V ($T_{\rm eff}=6400$ K). This
suggests a bolometric luminosity of only $L\sim$\,15 L$_\odot$, and hence a
difference in the distance modulus of $\Delta(m-M)\sim$\,10 mag, which
corresponds to a distance of $d\sim$\,500 pc, or a distance above the galactic
plane of $z\sim$\,250 pc. From a comparison of the acquisition image of
October 2002 with the imaging in December 1995, no proper motion could be
detected in excess of $\mu\sim0.02^{\prime\prime}$ yr$^{-1}$, but the proper
motion for a star at $d=500$ pc and galactic longitude of $l_{\rm
II}=277^\circ$ is also not expected to exceed this upper limit.

\label{lastpage}
\end{document}